\documentclass{q-DM-jetpl-1612.02326}   

\twocolumn  
\lat

\title{Dark matter from dark energy in $q$-theory}

\rtitle{Dark matter from dark energy in $q$-theory}

\sodtitle{Dark matter from dark energy in $q$-theory}

\author{F.R. Klinkhamer$^{+}$%
, G.E. Volovik$^{*\&}$
\thanks{e-mail: frans.klinkhamer@kit.edu\,;\,volovik@ltl.tkk.fi}
}

\rauthor{F.\,R.\,Klinkhamer, G.\,E.\,Volovik}

\sodauthor{Klinkhamer, Volovik}

\address{
$^+$Institute for
Theoretical Physics, Karlsruhe Institute of
Technology (KIT), 76128 Karlsruhe, Germany\\~\\
$^*$Low Temperature Laboratory, Aalto University,
P.O. Box 15100, FI-00076 Aalto, Finland\\~\\
$^\&$ Landau Institute for Theoretical Physics,
Russian Academy of Sciences, Kosygina 2, 119334 Moscow, Russia
}

\abstract{
A constant (spacetime-independent) $q$-field
may play a crucial role for the cancellation
of Planck-scale contributions to the
gravitating vacuum energy density.
We now show that a small spacetime-dependent
perturbation of the equilibrium $q$-field behaves
gravitationally as a pressureless perfect fluid.
This makes the fluctuating part of the
$q$-field a candidate for the inferred
dark-matter component of the present universe.
For a Planck-scale
oscillation frequency of the $q$-field perturbation,
the implication would be
that direct searches for dark-matter particles
would remain unsuccessful in the foreseeable future.
}

\PACS{95.35.+d, 95.36.+x, 98.80.Es, 98.80.Jk}

\begin{document}
\maketitle

\section{Introduction}
\label{sec:Introduction}

A condensed-matter-type approach to the
cosmological constant problem~\cite{Weinberg1988}
is given by $q$-theory~\cite{KV2008a,
KV2008b,KV2016-q-brane,KV2016-Lambda-cancellation}.
The aim of the $q$-theory formalism is to describe
the thermodynamics and dynamics of the deep quantum vacuum
without detailed knowledge of the microscopic (Planck-scale)
degrees of freedom.
Instead, an effective theory is considered with
one or more conserved $q$-fields.
For constant (spacetime-independent) $q$-fields,
the thermodynamics leads to an exact cancellation of the
quantum-field zero-point-energies in equilibrium, which
partly solves the cosmological constant problem.

It was already noted in Ref.~\cite{KV2008b}
that a rapidly-oscillating $q$-field could give a significant contribution
to the inferred dark-matter component of our present universe.
Here, we expand on this dark-matter aspect of
$q$-theory.

\section{Setup}
\label{sec:Setup}

Consider the particular realization of $q$-theory
based on a 3-form gauge field $A$ with a
corresponding 4-form field strength $F \propto q$
(see Refs.~\cite{KV2008a,KV2008b} and further references therein).
The action is now taken to include a kinetic term for
the $q$-field~\cite{KV2016-q-ball},
\vspace*{-1mm}
\begin{subequations}\label{eq:action-qdefinition12}
 \begin{eqnarray}
S&=&- \int_{\mathbb{R}^4}
\,d^4x\, \sqrt{-g}\,\left(\frac{R}{16\pi G_{N}} +\epsilon(q)
\right.
\nonumber\\[1mm]
&&
\left.
+\frac{1}{8}\,K(q)\, g^{\alpha\beta}\, \nabla_\alpha (q^2) \nabla_\beta (q^2) +\mathcal{L}^\text{\,SM}\right),
\label{eq:action}
\\[1mm]
q^2 &\equiv&- \frac{1}{24}\,
F_{\alpha\beta\gamma\delta}\,F^{\alpha\beta\gamma\delta}\,,\quad
F_{\alpha\beta\gamma\delta}\equiv
\nabla_{[\alpha}A_{\beta\gamma\delta]}\,,
\label{eq:qdefinition1}
\end{eqnarray}
\newpage  
\begin{eqnarray}
F_{\alpha\beta\gamma\delta}&=&q\sqrt{-g} \,\epsilon_{\alpha\beta\gamma\delta}\,,
\quad
F^{\alpha\beta\gamma\delta}=q \,\epsilon^{\alpha\beta\gamma\delta}/\sqrt{-g}\,, \quad
\label{eq:qdefinition2}
\end{eqnarray}
\end{subequations}
where the functions $\epsilon(q)$ and $K(q)$ in \eqref{eq:action}
involve only even powers of $q$, as $q$
is a pseudoscalar according to \eqref{eq:qdefinition2}
with the Levi--Civita symbol $\epsilon_{\alpha\beta\gamma\delta}$.
In the 4-form realization, the mass dimension of $q$ is 2.
Here, and elsewhere, we use natural units with $c=\hbar=1$
and take the metric signature $(-+++)$. For the curvature
tensors, we use the same conventions as in Ref.~\cite{KV2008b}.

The Lagrange density $\mathcal{L}^\text{\,SM}$
in the action \eqref{eq:action}
involves the fields of the standard model (SM) of elementary particle physics. In principle, it is also possible to replace
Newton's gravitational constant $G_{N}$ by a function $G(q)$,
but we will not do so in the present article as we already have explicit
$q$ derivatives in the action. Note that the energy density
$\epsilon(q)$ in the integrand of \eqref{eq:action}
may contain a constant term $\Lambda_\text{bare}$ of arbitrary sign.

With the definition
\begin{equation}
\label{eq:Kbar-def}
C(q )\equiv K(q)\,q^2 \,,
\end{equation}
the equations of motion for the
3-form gauge field can be written as
a generalized Maxwell equation,
\begin{align}
 \label{eq:gen-Maxwell-eq}  
 \nabla_\beta \left( \frac{d\epsilon(q)}{dq}
 -\frac12 \, \frac{d C(q)}{dq}\, \nabla_\alpha \, q\, \nabla^\alpha q
 - C(q) \, \Box \, q \right) = 0 \,.
\end{align}
The solution of this generalized Maxwell equation
is given by
\begin{align}
 \label{eq:gen-Maxwell-eq-solution-mu}  
  \frac{d\epsilon(q)}{dq}
  -\frac12 \, \frac{d C(q)}{dq}\, \nabla_\alpha \, q\, \nabla^\alpha q - C(q) \, \Box \, q  = \mu \,.
\end{align}
with an integration constant $\mu$.

The Einstein equation from \eqref{eq:action} reads
\begin{equation}
\label{eq:Einstein-eq}
R_{\alpha\beta} - \frac12\, g_{\alpha\beta}\,R = - 8\pi G_{N}
\left( T_{\alpha\beta}^{\,(q)} + T_{\alpha\beta}^\text{\,(SM)} \right) \,.
\end{equation}
The contribution of the 3-form gauge field to the energy-momentum tensor is
given by
\begin{eqnarray}
 \label{eq:energy-momentum-tensor-q}
  T_{\alpha\beta}^{\,(q)}
  &=&-\, g_{\alpha\beta} \left( \epsilon(q) - \mu \, q + \frac12 C(q) \, \nabla_\alpha \, q \, \nabla^\alpha q  \right)
 \nonumber\\[1mm]
&&
  + C(q) \nabla_\alpha \, q \, \nabla_\beta \, q \,,
\end{eqnarray}
where the solution \eqref{eq:gen-Maxwell-eq-solution-mu}
with integration constant $\mu$ has been used to simplify the expression.

Three remarks on the $q$-field energy-momentum tensor are in order.
First, for constant (spacetime-independent) $q$-fields, the
energy-momentum tensor \eqref{eq:energy-momentum-tensor-q}
has a cosmological-constant-type
term $-\,g_{\alpha\beta}\,\Lambda_\text{eff}(q)$  
with a gravitating vacuum energy density $\rho_{V}(q)$
different from energy density $\epsilon(q)$ of the action,
\begin{equation}
\label{eq:rho-V}   
\Lambda_\text{eff}(q)\,\big|_{q=\text{const.}}
=
\rho_{V}(q)
\equiv  
\epsilon(q) -  q\,\frac{d\epsilon(q)}{d q} \,,
\end{equation}
where $\mu$ has been replaced by $d\epsilon/dq$
according to \eqref{eq:gen-Maxwell-eq-solution-mu}.

Second, for nonconstant $q$-fields, we observe that terms with
$(dC/dq)\, (\nabla\, q)^2$ and $C\, \Box\, q$
have been absorbed completely by the constant $\mu$
in \eqref{eq:energy-momentum-tensor-q} and that the two remaining
terms with $C\, (\nabla\, q)^2$
have the same structure as if $q$ were a fundamental (pseudo-)scalar.

Third, in a cosmological context with $q=q(t)$,
the $q$-derivative terms in the action \eqref{eq:action}
allow the vacuum energy density $\rho_{V}(q)$ to change with
cosmic time $t$
even for constant gravitational coupling, $G(q)=G_{N}$
(compare with the discussion in the second paragraph
of Sec.~II in Ref.~\cite{KV2008b}).
For the moment, we postpone the study of these relaxation effects.

\section{Equilibrium q-field}
\label{sec:Equilibrium-q-field}

In equilibrium, the cosmological constant $\Lambda_\text{bare}$
(from quantum-field zero-point-energies, cosmological
phase transitions, or other origins)
is cancelled by a spacetime-independent
$q$-field of appropriate magnitude $q_{0}$.
This cancellation mechanism by the composite pseudoscalar $q$-field
is essentially different from a possible cancellation mechanism by a
fundamental pseudoscalar field $\phi$; see, in particular, the discussion of Sec.~2 in Ref.~\cite{KV2016-q-brane}.

Specifically, we have for the equilibrium state
\begin{subequations}\label{eq:q-equilibrium}
\begin{eqnarray}\label{eq:q-equilibrium-q0}
q(x)&=&\text{constant}=q_{0}\,,
\\[2mm]\label{eq:q-equilibrium-mu0}
\mu&=&\mu_{0}=d\epsilon/dq\,\big|_{q=q_{0}}\,,
\\[2mm]\label{eq:q-equilibrium-rV0}
\epsilon(q_{0})-\mu_{0}\, q_{0}&=&0\,.
\end{eqnarray}
\end{subequations}
According to \eqref{eq:energy-momentum-tensor-q}
and \eqref{eq:rho-V} for
constant $q$, the last equality \eqref{eq:q-equilibrium-rV0}
corresponds to the nullification of the gravitating vacuum energy
density~\cite{KV2008a},
\begin{equation}
\label{eq:rV-nullification}
\Lambda_\text{eff}(q_{0})
= 
\rho_{V}(q_{0})=0\,.
\end{equation}
A further stability condition is given by the positivity
of the inverse isothermal vacuum compressibility~\cite{KV2008a},
\begin{equation}
\big(\chi_{0}\big)^{-1} \equiv
\Bigg[\, q^2\;\frac{d^2  \epsilon(q)}{d  q^2}\,\Bigg]_{q=q_{0}} > 0\,.
\label{eq:chi0}
\end{equation}

Without additional matter,
the Einstein equation \eqref{eq:Einstein-eq} for the
equilibrium $q$-field \eqref{eq:q-equilibrium}
gives Minkowski spacetime with the metric
\begin{equation}
\label{eq:Minkowski-metric} 
g_{\alpha\beta}(x)\,\big|_\text{equil.}
=\eta_{\alpha\beta}=
\big[\ensuremath{\mathrm{diag}}( -1,\,  1,\, 1,\, 1)\big]_{\alpha\beta}\,,
\end{equation}
for standard Cartesian coordinates
$(x^{0},\,  x^{1},\,  x^{2},\, x^{3})= (t,\,  x,\,  y,\, z)$.

The question is still what the orders of magnitude are
of $q_{0}$ and $1/\chi_{0}$. If we assume that the
theory \eqref{eq:action} without the SM term
only contains a single energy scale, then that scale
must be of order of the Planck energy
\begin{equation}
\label{eq:E-Planck}
E_P \equiv (G_{N})^{-1/2}
     \approx 1.22 \times 10^{19}\,\text{GeV}\,.
\end{equation}
In that case, we have
\begin{subequations}\label{eq:q0-chi0inverse-scale}
\begin{eqnarray}
q_{0} \stackrel{?}{\sim} (E_P)^2\,,
\\[2mm]
1/\chi_{0} \stackrel{?}{\sim} (E_P)^4\,,
\end{eqnarray}
\end{subequations}
where the question marks are to remind us that these
estimates are based on an assumption.
For definiteness, we will phrase the rest
of the discussion in terms of the Planck-scale estimates
\eqref{eq:q0-chi0inverse-scale}, but the
discussion can be readily adapted if the relevant energy
scale is significantly different from the
Planck energy scale $E_P$.

The outstanding dynamical question is how the
equilibrium state is reached. It appears that energy
exchange between the vacuum $q$-field and the
matter fields plays a crucial role. For the special
case of massless-particle production, it has been
found that the equilibrium value $q_{0}$ is reached
dynamically and that
the final metric corresponds to the one of
Minkowski spacetime~\cite{KV2016-Lambda-cancellation}.
As the present article is explorative,
we will simply start from the equilibrium state.

\section{Pressureless perfect fluid}
\label{sec:Pressureless-perfect-fluid}

Now consider a small spacetime-dependent change of the
equilibrium $q$-field,
\begin{equation}\label{eq:q-perturbation-xi}
q(x)=q_{0}+ q_{0}\,\xi(x)\,,
\end{equation}
in terms of a dimensionless real scalar field $\xi(x)$
with $|\xi(x)|\ll 1$.  Also take
\begin{subequations}\label{eq:K-constant-q0-positivity}
\begin{eqnarray}\label{eq:K-constant}
K(q)&=&\text{constant}=(q_{0})^{-3}\,,
\\[2mm]
q_{0}&>& 0\,,
\label{eq:q0-positivity}
\end{eqnarray}
\end{subequations}
so that $C(q_{0})$ from \eqref{eq:Kbar-def} is positive.

The reduced Maxwell equation \eqref{eq:gen-Maxwell-eq-solution-mu}
for $\mu=\mu_{0}$  gives the following Klein--Gordon
equation~\cite{KV2016-q-ball}:
\begin{equation}
\label{eq:Klein-Gordon-eq-xi}
\Box\,\xi - \frac{1}{q_{0}}\,\Bigg[\, q^2\;\frac{d^2  \epsilon(q)}{d  q^2}\,\Bigg]_{q=q_{0}}\xi=0 \,,
\end{equation}
where higher-order $\xi$ terms have been omitted.
Neglecting, at first, the spatial derivatives of $\xi$,
the solution of \eqref{eq:Klein-Gordon-eq-xi}
is a rapidly-oscillating homogeneous function,
\begin{subequations}\label{eq:oscillating-xi-solution}
\begin{eqnarray}
\label{eq:oscillating-xi-solution-xi}
\xi(t) &=& a_{\xi}\,\sin\big(\omega\, t+ \varphi_{\xi}\big)\,,
\\[2mm]
\label{eq:oscillating-xi-solution-omega2}
\omega^2 &=& (q_{0})^{-1}  (\chi_{0})^{-1}
         \stackrel{?}{\sim} (E_P)^2\,,
\end{eqnarray}
\end{subequations}
where the small amplitude $a_{\xi}$ and the phase
$\varphi_{\xi}$ in \eqref{eq:oscillating-xi-solution-xi} are
determined by the boundary conditions
and where
the last estimate in \eqref{eq:oscillating-xi-solution-omega2}
follows from \eqref{eq:q0-chi0inverse-scale}.

For a time-dependent homogeneous perturbation $\xi(t)$
in \eqref{eq:q-perturbation-xi},
the energy-momentum tensor \eqref{eq:energy-momentum-tensor-q} becomes
\begin{subequations}\label{eq:T00andT11}
\begin{eqnarray}
T_{00}^{\,(q)}= \frac12\, q_{0}\, (\partial_t\, \xi)^2
               +\frac12\, (\chi_{0})^{-1} \xi^2\,,
\\[2mm]
T_{11}^{\,(q)}= \frac12\, q_{0}\, (\partial_t\, \xi)^2
               -\frac12\, (\chi_{0})^{-1} \xi^2\,.
\end{eqnarray}
\end{subequations}
Note that the structure of \eqref{eq:T00andT11} is the same
as that of a fundamental scalar field $\phi(t)$, which agrees
with the second remark at the end of Sec.~\ref{sec:Setup}.
With the rapidly-oscillating solution \eqref{eq:oscillating-xi-solution},
we get
\begin{subequations}\label{eq:energy-momentum-tensor-q-oscillating-xi}
\begin{eqnarray}
\label{eq:energy-momentum-tensor-q-oscillating-xi-00}
T_{00}^{\,(q)}&=& \frac12\, (\chi_{0})^{-1} \,(a_{\xi})^{2}\,,
\\[2mm]
\label{eq:energy-momentum-tensor-q-oscillating-xi-11}   
\langle T_{11}^{\,(q)}\rangle &=& \;\langle T_{22}^{\,(q)}\rangle  
\; =\;\langle T_{33}^{\,(q)}\rangle \;
\sim\;0\,, 
\\[2mm]
T_{01}^{\,(q)} &=& T_{02}^{\,(q)} =T_{03}^{\,(q)}= 0\,,
\end{eqnarray}
\end{subequations}
where the brackets $\langle\ldots \rangle$    
in \eqref{eq:energy-momentum-tensor-q-oscillating-xi-11}
denote the average over a large time interval $T \gg \pi / \omega$
and where  we have also added the results for the other components.

At this moment, recall that a perfect fluid has an
energy-momentum tensor of the form
\begin{equation}
\label{eq:energy-momentum-tensor-perfect-fluid}
T_{\alpha\beta} = P\,g_{\alpha\beta}+(\rho+P)\,U_{\alpha}U_{\beta}\,,
\end{equation}
with $g_{\alpha\beta}=\eta_{\alpha\beta}$ and
$U^{\alpha}=(1,\, 0,\, 0,\, 0)$ in Minkowski spacetime.
From \eqref{eq:energy-momentum-tensor-q-oscillating-xi},
we then conclude that
the homogeneous $\xi$ perturbation of the vacuum $q$-field
behaves as a perfect fluid with the following values for the
energy density and pressure:
\begin{subequations}\label{eq:rho-P-fluid}
\begin{eqnarray}
\rho^{\,(q-\text{perturbation})}&=&
\frac12\, (\chi_{0})^{-1} \,(a_{\xi})^{2}\,,
\\[2mm]
P^{\,(q-\text{perturbation})}
&
\sim  
&0\,,
\end{eqnarray}
\end{subequations}
where $a_{\xi}$ with $|a_{\xi}|\ll 1$
is determined by the initial boundary conditions
[in a cosmological context, taken at the moment when the
homogeneous vacuum energy density $\rho_{V}(t)$ has reached
its final near-zero value; see Sec.~5~for further discussion].

Next, consider additional space-dependence of the
$\xi$ field with a typical length-scale
\begin{equation}
\label{eq:L}
L \gg c/\omega \stackrel{?}{\sim} \hbar\, c/E_P \sim 10^{-35}\,\text{m}\,,
\end{equation}
where $\hbar$ and $c$ have been temporarily reinstated
and the estimate \eqref{eq:oscillating-xi-solution-omega2}
for $\omega$ has been used.
(For applications
in a cosmological context, the length-scale $L$ must be less than the cosmological length-scale $c/H_{0} \sim 10^{26}\,\text{m}$.)
The corrections to
\eqref{eq:energy-momentum-tensor-q-oscillating-xi}
will then be small, namely of order $q_{0}\,\xi^2/L^2$,
which is a factor $(L\,\omega)^{-2}\ll 1$
times the leading term
\eqref{eq:energy-momentum-tensor-q-oscillating-xi-00}.
For such large-scale perturbations, there will be,
to high precision, a pressureless perfect fluid and
this fluid will cluster gravitationally,
just as cold dark matter with standard Newtonian gravitation and dynamics.

\section{Conclusion}
\label{sec:Conclusion}

In the present article, we have shown that a small
perturbation of the equilibrium $q$-field behaves
gravitationally as a pressureless perfect fluid.
As such, the fluctuating part of the $q$-field is
a candidate for the inferred cold-dark-matter component    
of the present universe (see, e.g., Sec.~26
of Ref.~\cite{PDG2016} for a review).

There are, however, many open questions.
The main question concerns the amount of dark matter (DM)
vs. that of dark energy (DE).
The presently observed universe is very close to equilibrium,
$\rho_{V,\,\text{obs}} \ll (E_P)^4$,
but still somewhat away from it as
$\rho_{V,\,\text{obs}} \ne 0$, which gives in our framework
$q_\text{obs}= q_{0}+ \delta q$  with  $0<|\delta q|\ll |q_{0}|$.
We, therefore, need to find a mechanism which results in
the small constant perturbation $\delta q$ of the $q$-field
for the present universe.
Several possible mechanisms have been explored,
including effects from $\text{TeV}$-mass particle
decays~\cite{ArkaniHamed-etal2000,KV2009-electroweak,%
K2011-field-theoretic-model} or
from neutrinos with sub-$\text{eV}$ masses~\cite{KV2011-review}.

A final, definitive calculation
of the appropriate $\delta q$  and the corresponding
\mbox{$\rho_{V}=-P_{V}>0$} (to be interpreted
as the inferred ``dark-energy''
component of the present universe) is not yet available.
The same can be said of a further perturbative component $\xi(x)$
in the $q$-field, $q(x)= q_{0}+ \delta q + q_{0}\,\xi(x)$,
which would determine the present amount of dark matter.
Experimentally, we have
$\rho_\text{DE}/\rho_\text{DM} \sim 3$
and $\rho_\text{DE}+\rho_\text{DM}\sim
\rho_\text{critical}\equiv 3H_{0}^2/8\pi G_{N}$ (see, e.g., Secs.~26 and 27
of Ref.~\cite{PDG2016}).

Another open question concerns the effects on the clustering
of subleading terms in the $q$-field energy-momentum tensor
coming from the spatial derivatives of the perturbation $\xi(x)$.
A further issue is the ultimate origin of
the required perturbations
in the dark-matter energy density with length scales
obeying \eqref{eq:L}, which may or may not involve
inflation-type processes.

To conclude,
even though we still need to work out many details,
we do have a clear prediction.
Our proposal is to identify dark matter with an oscillating
component of the $q$-field, most likely having a Planck-scale frequency.
If correct, this implies that direct detection of dark-matter
particles will fail, at least in the foreseeable future with the
currently available energies.

We thank T. Mistele and M. Savelainen for help in obtaining
Eq.~\eqref{eq:energy-momentum-tensor-q}.
The work of GEV has been supported by the European Research Council
(ERC) under the European Union's Horizon 2020 research and innovation programme (Grant Agreement No. 694248).


\begin{thebibliography}{99}

\bibitem{Weinberg1988}
S. Weinberg,
``The cosmological constant problem,''
Rev. Mod. Phys.  {\bf 61}, 1 (1989).


\bibitem{KV2008a}
F.R. Klinkhamer and G.E. Volovik,
``Self-tuning vacuum variable and cosmological constant,''
Phys. Rev. D \textbf{77}, 085015 (2008), arXiv:0711.3170.

\bibitem{KV2008b}
F.R. Klinkhamer and G.E. Volovik,
``Dynamic vacuum variable and equilibrium approach in cosmology,''
Phys. Rev. D \textbf{78}, 063528 (2008), arXiv:0806.2805.

\newpage
\bibitem{KV2016-q-brane}
F.R. Klinkhamer and G.E. Volovik,
``Brane realization of $q$-theory and the cosmological constant problem,''
JETP Lett.\  {\bf 103}, 627 (2016),
arXiv:1604.06060.  

\bibitem{KV2016-Lambda-cancellation}
F.R. Klinkhamer and G.E. Volovik,
``Dynamic cancellation of a cosmological constant and
approach to the Minkowski vacuum,''
Mod.\ Phys.\ Lett.\ A {\bf 31}, 1650160 (2016),
arXiv:1601.00601. 

\bibitem{KV2016-q-ball}
F.R.~Klinkhamer and G.E.~Volovik,
``Propagating $q$-field and $q$-ball solution,''
arXiv:1609.03533.  



\bibitem{PDG2016}
C. Patrignani \textit{et al.} (Particle Data Group),
``The review of particle physics (2016),''
Chin. Phys. C {\bf 40}, 100001 (2016)          
[available from \texttt{http://pdg.lbl.gov/}].


\bibitem{ArkaniHamed-etal2000}
N. Arkani-Hamed, L.J. Hall, C. Kolda, and H. Murayama,
``New perspective on cosmic coincidence problems,''
Phys.\ Rev.\ Lett.\  {\bf 85}, 4434 (2000),
arXiv:astro-ph/0005111.

\bibitem{KV2009-electroweak}
F.R. Klinkhamer and G.E. Volovik,
 ``Vacuum energy density kicked by the electroweak crossover,''
Phys. Rev. D {\bf 80}, 083001 (2009),
arXiv:0905.1919.

\bibitem{K2011-field-theoretic-model}
F.R.~Klinkhamer,
``Effective cosmological constant from TeV-scale physics: Simple field-theoretic model,''
Phys.\ Rev.\ D {\bf 84}, 023011 (2011),
arXiv:1101.1281.  


\bibitem{KV2011-review}
F.R.~Klinkhamer and G.E.~Volovik,
``Dynamics of the quantum vacuum: Cosmology as relaxation to the equilibrium state,''
J.\ Phys.\ Conf.\ Ser.\  {\bf 314}, 012004 (2011),
arXiv:1102.3152.  



\end{thebibliography}
\end{document}